\documentstyle[aps,epsf,multicol,prl]{revtex}
\begin{document}
\draft

\author{J. Reichert$^1$, H. B. Weber$^{1,*}$, M. Mayor$^{1,*}$, and H. v. L\"ohneysen$^{2,3}$}
\address{$^1$Forschungszentrum Karlsruhe, Institut f\"ur Nanotechnologie, D-76021 Karlsruhe}
\address{$^2$Forschungszentrum Karlsruhe, Institut f\"ur Festk\"orperphysik, D-76021 Karlsruhe}
\address{$^3$Physikalisches Institut, Universit\"at Karlsruhe, D-76128 Karlsruhe}
\sloppy
\title{Low-Temperature Conductance Measurements On Single Molecules}
\maketitle
\begin{abstract}
An experimental protocol which allows to perform conductance spectroscopy on organic molecules at low temperatures ($T \approx 30$~K) has been developed. This extends the method of mechanically controlled break junctions which has recently demonstrated to be suitable to contact single molecules at room temperature. The conductance data obtained at low $T$ with a conjugated sample molecule show a highly improved data quality with a higher stability, narrower linewidth and substantially reduced noise.
Thus the comparability of experimental data with other measurements as well as with theoretical simulations is considerably improved.
\end{abstract}
\pacs{}

\begin{multicols}{2}
\narrowtext

Molecular electronics employing single molecules as active functional units is a promising new technological concept of fast-growing
interest. With molecule-based systems it may, for example, be possible to achieve higher storage densities in conjunction with a simplified production technology \cite{2002 Heath}. Our research is dedicated to understand the fundamental processes of electron conduction through individual molecules. This is a crucial requirement for the purposeful design of molecules for electronic functionalities.
Here we present a method to perform low-temperature conductance measurements (T\( \approx  \)30~K)
of single-molecule contacts with the mechanically controlled break-junction (MCB) technique. While only few such measurements exist at room temperature (RT)~\cite{1997 reed,1999 Kergueris,2002 Reichert,2002 Mayor}, low-$T$ measurements are desirable for several reasons: First, the physics
is then considerably simpler (e.g. no vibrations). Further, at low $T$ it should be possible to reduce atomic-scale rearrangements on the electrode surface
which strongly affect the measurements at RT~\cite{2002 Weber}. Another complementary method has also been used for single-molecule contacts at low $T$~\cite{2002 Park}. In contrast to the MCB technique discussed here, this electromigration  method allows for a field-effect transistor setup, but does not provide well-controlled distance control between the contacts. 

\begin{figure}[tb]
\epsfxsize5cm
\centering\leavevmode\epsfbox{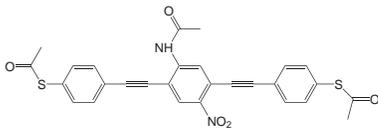}
\caption{ Investigated molecular rod 2,5-di(2'-(para-acetylmercaptophenyl)ethinyl)-4-nitro-acetylaniline, with acetyl- protected terminal thiol groups for immobilization on the gold surfaces of the MCB.}
\label{fig: mol}
\end{figure}

Earlier experiments have demonstrated that the conductance of single-molecule junctions can be determined at RT by means
of MCBs~\cite{2002 Reichert}, which provide a tunable-distance electrode pair. It consists of a metallic wire structured on a flexible substrate with a narrow bridge-like constriction~\cite{1992 Muller,1996 Ruitenbeek}, serving as a  predetermined fracture point. If the substrate is bent, the bridge on top stretches, until it finally tears apart, yielding a nanoscopic pair of electrodes facing each other. Due to the ultra-low transmission ratio between the inflection of the substrate and the expansion at the surface the electrode distance is controllable with sub-\AA ngstrom precision and possesses a high mechanical stability. However, cooling down a given contact is barely possible because of the different thermal expansion coefficients of the materials used which result in a badly controlled drift of the electrode spacing ($\approx 1$ nm in our experiments upon cooling from RT to 30 K). An electronic adjustment (feedback control)
of the electrode gap during cooling would require the knowledge
of the absolute distance during this process which is not easily
accessible experimentally.
The molecules investigated in this experiment are rigid rod-like organic
molecules with a length of approximately 2 nm (Fig.~\ref{fig: mol}). The
molecules possess a \( \pi  \)-electron system between the acetyl-protected
sulfur functional groups at their ends, which are designed to split off when the molecules bind covalently to the gold electrode surface. This procedure, however, is expected to work only at RT, because at low $T$ there is not enough energy available for the deprotection reaction. 
Hence, we have to find an experimental protocol which yields low-$T$ contacts, but avoids temperature sweeps of a given junction as well as surface reactions at low $T$.

\begin{figure}
\epsfxsize4.5cm
\centering\epsfbox{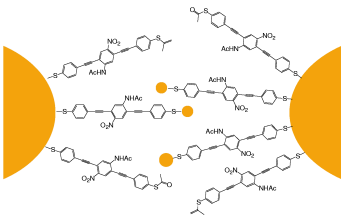}
\caption{Scenario between the MCB after closing and reopening at RT. Upon contact with the opposite electrode, the protection groups of the formerly bridging molecules have partly been replaced by clusters of a few Au atoms.}
\label{fig: electrodes}
\end{figure}

This protocol starts with the procedure for RT measurements~\cite{2002 Reichert}.
When applying a solution of acetate-protected thiol-functionalized
molecules (Fig.~\ref{fig: mol}) on a freshly opened pair of electrodes,
a molecular layer on the gold surface is formed. The molecules split
off their acetate protection group upon contact and bind covalently
with one end to the gold surface. At this point, the other side of
the molecule remains protected~\cite{1999 Gryko}. Due to the short
application times ($\approx 60$~s) we do not expect a cohesive monolayer
in our experiments~\cite{1995 Tour}. The high surface mobility of
the polarisable molecules allows to draw them into the contact
region between the electrodes by means of a dc electric field which is provided by a finite bias voltage. Hence we decrease
the electrode distance while $U = 1-1.5$ V is applied and we monitor the current until the first molecule touches the opposite electrode, splits off its protection group and binds covalently to the counter electrode. At this moment
a sharp increase in current is recorded. Often, a molecular junction is now established which allows to record stable current-voltage characteristics (IVs) in a voltage window of [-1.2~V,1.2~V] for the given molecules. More details about the RT measurements can be found in~\cite{2002 Reichert}. 
It is expected that all bonds in the molecule are more stable than a Au-Au bond, an assumption which has been confirmed by molecular dynamics simulations~\cite{2002 Kruger}. There, it has been shown that when a thiol-ended molecule is pulled off the gold surface, the molecule rather picks one or more Au atoms out of the surface than breaking the Au-S bond. 
Consequently, the new protocol is based on the idea of terminating the molecules in the junction with a Au atom (or possibly a few Au atoms) by closing and reopening the junction (see Fig.~\ref{fig: electrodes}). Then, the junction can be cooled down while it is open and reestablished at low $T$ without involving any further organic chemical reactions (the formation of a Au-Au bond is considered to have a negligible activation barrier).

\begin{figure}[t]
\epsfxsize6cm
\centering\epsfbox{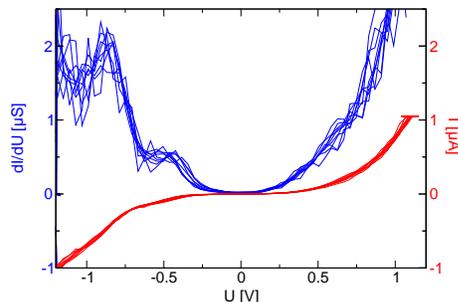}
\caption{Current-voltage raw data of a molecular junction
at room temperature. Red: the current as a function of the bias voltage
$U$. Blue: numerically differentiated data $dI/dU$ (after~\protect\cite{2002 Reichert}).}
\label{fig: asym-RT}
\end{figure}

The experiment was made in a UHV-compatible chamber. The break-junction setup was mounted on a continuous-gas-flow cryostat inset which allows to cool down the sample to  $\approx 30$ K$\pm 10$ K (The large error bar is due to insufficient thermal equilibrium in our setup where rather quick measurements are required. We did not perform {\it in situ} thermometry on the chip, the difference between the sample's temperature and that of the thermometer was determined in a separate calibration experiment).
After applying the molecules from solution the junction was established in vacuum at RT until a resistance of the order of 10 k$\Omega$ was observed. This corresponds to numerous molecules which close the gap. Then the usual procedure was pursued until a single-molecule signature was found. After that the junction was reopened and then cooled down to $30$ K within $\approx 15$ minutes. It is crucial to pump the chamber thoroughly to the best available base pressure (in our case 3$\cdot 10^{-8}$ mbar) prior to cooling down. Otherwise, adsorbates will condense on the electrode surface and will prevent the reestablishment of a molecular junction at low $T$. In addition to turbomolecular pumps and ion getter pumps, we further used a cup-like shield at cryogenic temperatures to provide a cold trap close to the electrode pair. 
At $T\approx 30$ K, the junction was reapproached. Similarly to RT, first a tunnel current and then a sudden increase of the current was observed. At this stage, stable IVs could be recorded which will be discussed below.
For comparison, we first show previously published data from RT measurements: Fig.~\ref{fig: asym-RT} shows nine IVs as well as the differential conductance $dI/dV$ of the molecule in a stable configuration. Rounded step-like features are visible in the IV (red lines) which appear as broadened
peaks in $dI/dV$ (blue lines). The peak structure can qualitatively be explained within a coherent transport scenario \cite{2002 Weber,2002 Heurich}: As soon as a molecular orbital gets in resonance with the Fermi level of one of the electrodes, transport through this molecular orbital may be enabled resulting in a step-like increase of the current. After observing a stable spectrum  for several minutes as shown in Fig.~\ref{fig: asym-RT}, the IV often spontaneously changes into a new characteristics which may be also reproducible for several minutes. 
This jump-like switching is attributed to microscopic atomic or molecular rearrangements in the junction which are typical for a single-molecule junction. Note also the considerable noise in the data which is not created by the measurement electronics but by the molecular junction itself, as verified in control experiments.\\ 

\begin{figure}
\epsfxsize6cm
\centering\epsfbox{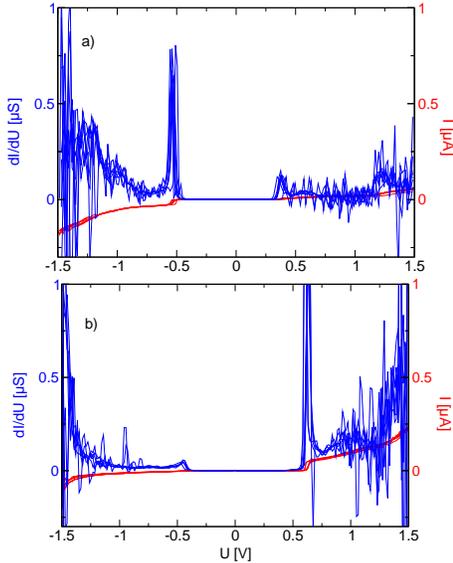}
\caption{a),b)Current-voltage raw data (red lines) and the
differential conductance $dU/dI$ (blue lines) as a function of the applied
bias voltage $U$ at 30~K. First a) was recorded. Then the junction was opened and closed again, resulting in the different configuration b).}
\label{fig: 30K-01}
\end{figure}

Low-$T$ data taken at $T = 30$ K  are shown in Fig.~\ref{fig: 30K-01}a. The low-bias region displays a strongly suppressed conductance ($\leq 1$nS). The linewidth of the peaks in $dI/dV$ is much smaller compared to RT measurements (of the order of 250~meV FWHM at RT, 30~meV at 30~K) and the overall data quality has substantially improved (less noise). Due to the narrower linewidth smaller peaks which cannot be seen at RT appear. For example, the peak at $U=-0.53$~V has a substantially reduced counterpart at opposite bias ($U=0.38$~V), which is not visible in the RT data.
The data observed for this spatially asymmetric molecule are always asymmetric with respect to voltage inversion, as discussed in detail in \cite{2002 Reichert}. This indicates that only very few molecules (most likely a single one) contribute to the conductance, because a large number of molecules, randomly oriented, would result in symmetric IVs. The conductance pattern at Fig.~\ref{fig: 30K-01}b was obtained only minutes after the pattern in Fig.~\ref{fig: 30K-01}a, after opening and closing the junction. The two spectra appear roughly as mirror images of each other with respect to voltage inversion. Apparently, another oppositely oriented molecule has replaced the one shown in Fig.~\ref{fig: 30K-01}a. 
While the absolute value of the current, taken at $U=1$ V, varies for different junctions from $0.1~\mu$A to $1~\mu$A at RT, the current in the present junction is $I \approx 60 -100$~nA for the two junctions shown here. Whether this reduction stems from different microscopic arrangements, from reduced molecular motion or from suppressed thermal activation can not be deduced from the data. We do not see signs that adsorbates influence the measurements (e.g. Ohmic behaviour at low voltages) as was observed in previous experiments with insufficient vacuum.
Another improvement of the low-$T$ measurements is the smaller range of different IVs. For the RT measurements a rather broad range of sample-to-sample fluctuations was attributed to different microscopic realizations \cite{2002 Reichert,1999 Yaliraki}. In particular microscopically different thiol-gold junctions 
were found to strongly affect the energetic level as well as the charge distribution of the HOMO. The difference of the bonding energies of these different realizations are, however, only small \cite{2002 Weber} and thus no configuration is clearly statistically preferred. The smaller number of variations in the recorded low-T spectra might indicate that here energetically favourable configurations occur more frequently. 
The presented method improves the investigation of single-molecule junctions, contributing to both, the comparability with theory as well as with other experiments. Finally, genuine low-$T$ experiments, for example with superconducting electrodes are now possible.

$^*$e-mail:heiko.weber@int.fzk.de; marcel.mayor@int.fzk.de

\end{multicols}
\end{document}